\documentclass{iaus}
\usepackage{graphicx}

\title[Dust distribution in circumstellar shells] 
{Dust distribution in circumstellar shells}

\author[van Marle et al.]   
{A.~J. van Marle$^1$,
  Z. Meliani$^1$, 
  R. Keppens$^1$,
 \and L. Decin$^2$}

\affiliation{$^1$Centre for Plasma Astrophysics, K.U.Leuven, Celestijnenlaan 200B, 3001 Leuven, Belgium  \\[\affilskip]
$^2$Institute of Astronomy , K.U.Leuven, Celestijnenlaan 200D, 3001 Leuven, Belgium }

\pubyear{2011}
\volume{IAU283}  
\pagerange{1--2}
\jname{Planetary nebulae, an Eye to the Future}
\editors{A.C. Editor, B.D. Editor \& C.E. Editor, eds.}
\begin{document}

\maketitle

\begin{abstract}
We present numerical simulations of the hydrodynamical interactions that produce circumstellar shells. 
These simulations include several scenarios, such as wind-wind interaction and wind-ISM collisions. 
In our calculations we have taken into account the presence of dust in the stellar wind. 
Our results show that, while small dust grains tend to be strongly coupled to the gas, large dust grains are only weakly coupled. 
As a result, the distribution of the large dust grains is not representative of the gas distribution. 
Combining these results with observations may give us a new way of validating hydrodynamical models of the circumstellar medium. 
\end{abstract}

\firstsection 
\section{Intoduction}
The winds of cool stars contain a large amount of dust. 
This dust is crucial in driving the wind of the star and, at larger distances, it moves along with the flow of the wind (\cite[Lamers \& Cassinelli 1999]{LC1999}). 
When the gas collides with the surrounding medium (interstellar medium, a previously ejected wind, wind from an other star, etc.) it slows down to form a shell, creating a velocity difference between gas and dust. 
Whether the dust will follow the change in velocity depends on the strength of the coupling between gas and dust, which in turn depends on such parameters as gas density, velocity difference, temperature and the size of the dust grains.  
The larger the dust grains, the more momentum they have and the more force will be required to slow them down. 
In this paper we use numerical models of a stellar wind colliding with its surrounding medium to investigate the behavior of the dust grains in the resulting shell.

\section{Model}
We use the {\tt mpi-amrvac} code (\cite[Keppens et al. 2011]{Ketal2011}) for our calculations, which solves the conservation equations of hydrodynamics on a fixed grid. 
For these simulations we also include the effect of optically thin radiative cooling. 
The presence of dust is included using a two-fluid approximation, where the dust is treated as a gas without internal pressure (\cite[van Marle et al. 2011]{vMetal:2011}), with the interaction between gas described as a drag force (\cite[Kwok 1975]{K1975}):
\begin{equation}
{\mathbf{f}}~\propto~n_{\rm d} \rho a_{\rm d}^2 \Delta\,{\mathbf{v}}\sqrt{\Delta\,{\mathbf{v}}^2 + v_{\rm t}^2}
\end{equation}
with $n_{\rm d}$ the dust particle density, $\rho$ the gas density, $a_{\rm d}$  the radius of the dust grains, $\Delta\,{\mathbf{v}}$ the velocity difference between gas and dust and $v_{\rm t}$ the thermal speed of the gas. 
We simulate the interaction between a stellar wind ($10^{-6}\,{\rm M}_\odot/{\rm yr}$, 15\,km/s) with the interstellar medium (ISM), which has a density of 2 particles per cm$^3$. 
The computational domain has a radius of 0.5\,pc and an opening angle of 180$^{\rm o}$. 
The adaptive mesh refinement provides a maximum effective resolution of 3200$\times$800 gridpoints.
We assume that the ISM contains no dust and that 0.5\% of the mass of the wind is in the form of dust grains. 
The dust grains are assumed to be made of carbon (internal density 2.9\,g/cm$^3$). 
We run the simulation twice, once with a grain radius of 0.005\,$\mu m$ and once with a grain radius of 0.05\,$\mu m$. 

\begin{figure}[t]
 \includegraphics[width=0.48\columnwidth]{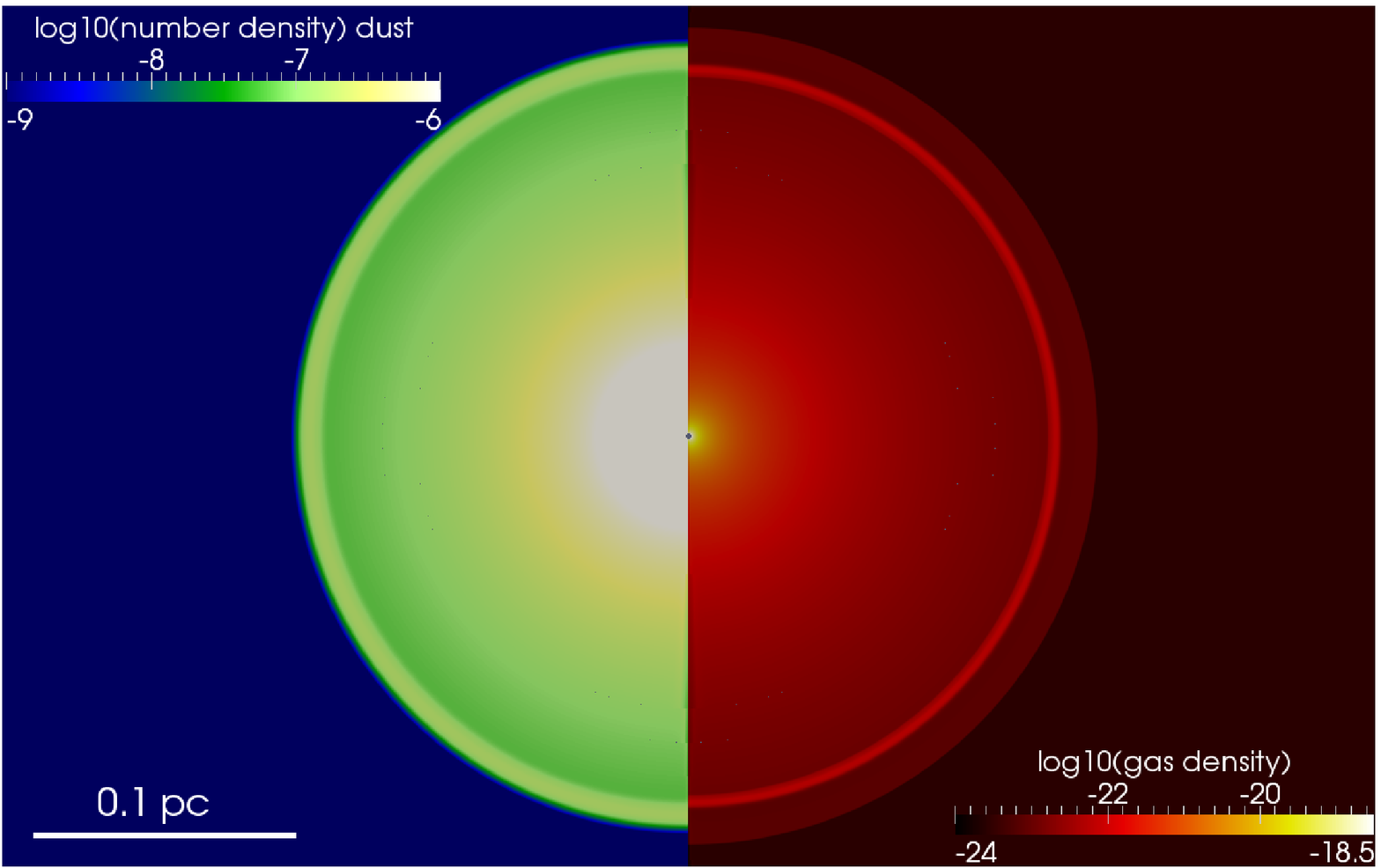}
 \includegraphics[width=0.48\columnwidth]{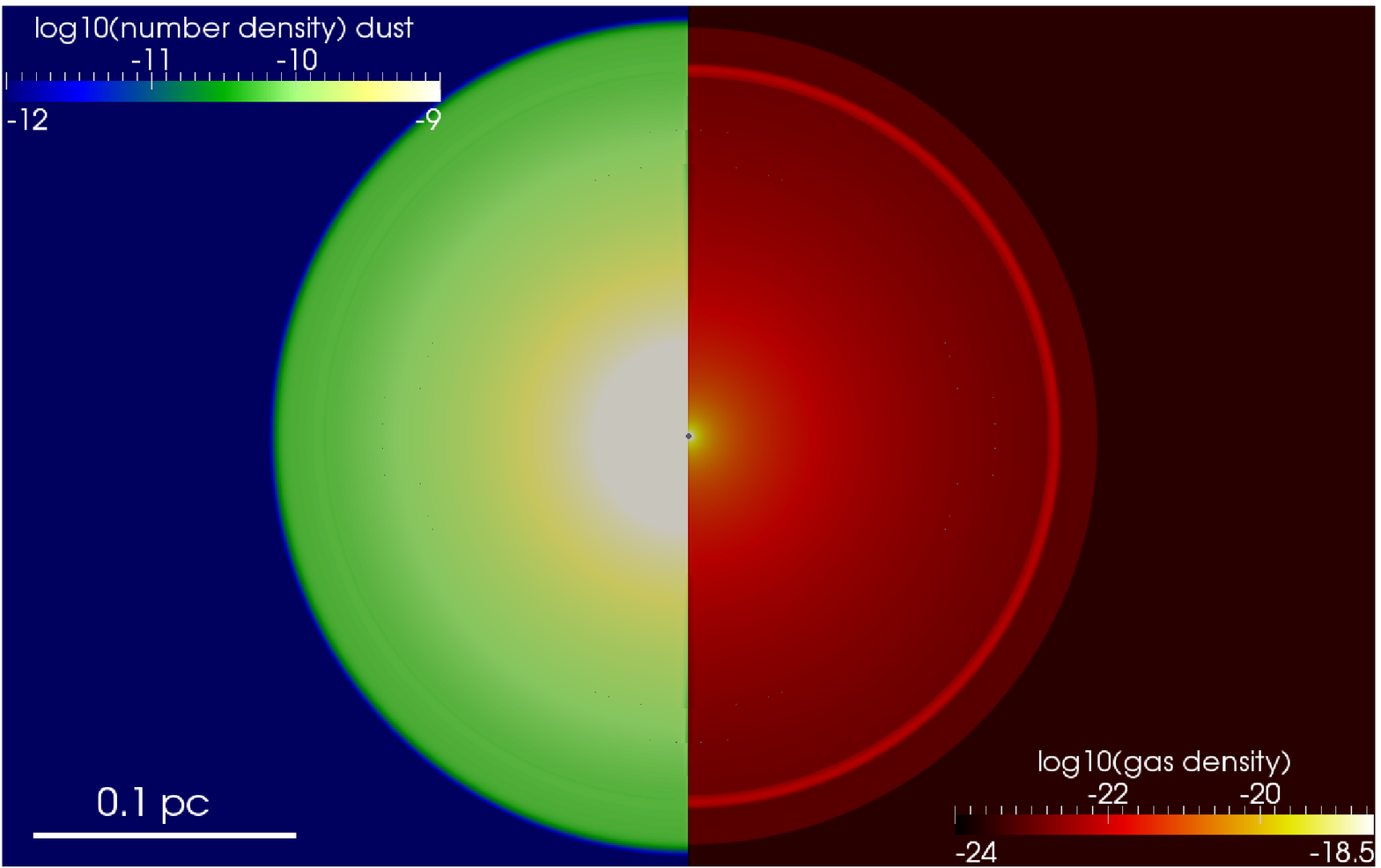}
 \caption{Gas mass density in g/cm$^3$ (right panels) and dust particle density (left panels) of the circumstellar medium after 10\,000 years. 
The small (0.005\,$\mu m$) dust grains (left figure)} show a shell at the contact discontinuity between shocked wind and shocked ISM, indicating that the drag force effectively couples them to the gas. 
The large (0.05\,$\mu m$) grains (right figure) show no shell, since their larger momentum makes them more difficult to slow down. 
   \label{fig:result}
\end{figure}

\section{Results}
Figure \ref{fig:result} shows the result of our simulations after 10\,000 years. 
As the gas hits the ISM, it creates a shell of shocked gas (shocked ISM on the outside and shocked wind material on the inside). 
The shocked wind, which has the highest density cools very effectively (raditive cooling depends on the gas density squared), causing the inner part of the shell to be compressed. 
At this stage the shell does not show instabilities. 
These will start to appear as the shell reaches a larger radius, where the density of the wind becomes less than the density of the interstellar medium it sweeps up, making the shell vulnerable to Rayleigh-Taylor instabilities. 
The small (0.005\,$\mu m$) dust grains show a shell-like structure around the contact discontinuity between shocked wind and shocked ISM, which proves that the drag force couples them to the gas and has effectively reduced their speed to the same velocity as the shocked gas. 
The larger (0.05\,$\mu m$) dust grains do not form a shell. 
Their momentum allows them to move almost freely through the shocked gas.

Whether dust grain distribution will follow the morphology of the circumstellar gas depends on the effectiveness of the coupling between dust and gas through the drag force. 
For small grains, this coupling is very effective, forcing the grains to follow the gas. 
For large grains, the coupling is too weak. 
It is possible to differentiate between grain sizes in observations. 
This will provide us with a new way of validating hydrodynamical simulations of the circumstellar environment. \\ 

{\small A.J.v.M.\ acknowledges support from FWO, grant G.0277.08 and K.U.Leuven GOA/09/009.}

\end{document}